\documentclass[aps,prd,showpacs,preprint,tightenlines]{revtex4}

\usepackage{graphicx}% Include figure files

\begin{document} 

% YY-NN = your latest CBX number 

\preprint{\vbox{\hbox{\hfil CLNS 02/1775}
                \hbox{\hfil CLEO 02-01}
}}

\title{
\vspace*{4cm}
Measurement of the Ratio of Branching Fractions \\
of
the $\Upsilon$(4S) to Charged and Neutral $B$ Mesons}

\date{\today}

\begin{abstract} 
The ratio of charged and neutral
$B$ meson production at the $\Upsilon$(4S),
$f_{+-}/f_{00}$, is measured through
the decays $\bar{B} \rightarrow D^{*}\ell^{-} \bar{\nu}_{\ell}$,
reconstructed using a partial reconstruction method
where the $D^*$ is detected only through a pion daughter from the
decay $D^*\rightarrow D\pi$.
Using data collected
by the CLEO II detector,
the charged and neutral $B$ decays are measured in such a way that their ratio is
independent of decay model, limited mainly by the uncertainty in
the relative efficiency for detecting neutral and charged
pions.
This measurement yields the ratio of production fractions times the ratio of
semileptonic branching fractions, $f_{+-}b_{+}\over f_{00}b_0$.
Assuming that $b_+\over b_0$ is equal to the lifetime ratio
$\tau_+\over\tau_0$ and using the world average value of $\tau_+\over\tau_0$ as input, we obtain
$f_{+-}/f_{00}=1.058\pm 0.084\pm 0.136$.

\end{abstract}

%Author list here
%\input cleoauthor.rev

\author{(CLEO Collaboration)}
\pacs{13.20.He, 13.65.+i }

\maketitle

\begin{center}
S.~B.~Athar,$^{1}$ P.~Avery,$^{1}$ H.~Stoeck,$^{1}$
J.~Yelton,$^{1}$
G.~Brandenburg,$^{2}$ A.~Ershov,$^{2}$ D.~Y.-J.~Kim,$^{2}$
R.~Wilson,$^{2}$
K.~Benslama,$^{3}$ B.~I.~Eisenstein,$^{3}$
% ...
\end{center}
 
\small
\begin{center}
$^{1}${University of Florida, Gainesville, Florida 32611}\\
$^{2}${Harvard University, Cambridge, Massachusetts 02138}\\
$^{3}${University of Illinois, Urbana-Champaign, Illinois 61801}\\
% ...
\end{center}
\newpage 

\section{Introduction}
The $\Upsilon$(4S) resonance decays predominantly to charged or neutral $B$ meson pairs. 
The cleanliness of exclusive pair production in
$e^+e^-\rightarrow\Upsilon$(4S)$\rightarrow B\bar B$ has made this mechanism a primary
vehicle for the study of the light pseudoscalar $B$ mesons.
Given the similar masses of the charged and neutral $B$, it is reasonable to expect that
the respective production fractions, $f_{+-}$ and $f_{00}$, are both around 50\%.
It has been noted, however, that mass and Coulomb effects could lead to corrections of as much as 20\% on the
ratio $f_{+-}/f_{00}$\cite{ftheory}.

Given the low efficiencies for $B$ reconstruction, it is not straightforward to measure
this ratio directly, and the uncertainty on its value is a major source of systematic
error for measurements of fundamental parameters at the $\Upsilon$(4S).
Currently, the most successful method involves measuring the rates at the $\Upsilon$(4S) of purely spectator decays of charged and neutral $B$ mesons to isospin related final states.
Each rate is proportional to the product of the branching fraction of the decay, $b_+$ or $b_0$, and $f_{+-}$ or $f_{00}$.
The ratio of the rates is equal to $b_+f_{+-}/b_0f_{00}$.
The ratio $b_+/b_0$ is equivalent to the ratio of lifetimes, $\tau_+/\tau_0$, if one
assumes that the partial decay widths are equal.  
Using an independently measured lifetime ratio, one may thus obtain
$f_{+-}/f_{00}$.  
We report here such a measurement of $f_{+-}/f_{00}$ via partial reconstruction of the
exclusive decays $B^- \rightarrow D^{*0} \ell^-\bar{\nu}_{\ell}$ and $\bar{B}^0 \rightarrow D^{*+}
\ell^- \bar{\nu}_{\ell}$.

The exclusive decay $\bar{B}^0
 \rightarrow D^{*+} \ell^- \bar{\nu}_{\ell} (D^{*+}\rightarrow D^0 \pi^+)$ has been previously
reconstructed in CLEO~II data
using a partial reconstruction technique\cite{artuso}. 
In this technique,
the $D^{*}$ is identified without reconstructing the $D$ 
meson, and the presence of the neutrino is inferred from conservation 
of momentum and energy.  
This method results in a gain of as much as a factor 20 in statistics 
compared to the full reconstruction method.
Partial reconstruction has been used by CLEO to tag neutral $B$'s
for measurements of the semileptonic branching 
fraction $b_0$\cite{artuso}, and the 
mixing parameter $\chi_d$\cite{mssthesis}.
The measurement we report here includes the first reconstruction of the exclusive decays 
$B \rightarrow D^* \ell^- \bar{\nu}_{\ell}(D^*\rightarrow D\pi^0)$ using the partial reconstruction method.  

\section{Data and Event Selection}

This analysis uses 2.73~fb$^{-1}$ of $e^+e^-$ 
annihilation data recorded at the $\Upsilon$(4S) resonance (on-resonance) and 1.43~fb$^{-1}$ taken at 60~MeV below the resonance (off-resonance).
These data were collected with the CLEO~II\cite{cleoii} detector at the Cornell Electron Storage Ring (CESR) between
1990 and 1995. 
The on-resonance sample includes 2.89 million 
$B\bar{B}$ events.  
Hadronic events are selected by requiring
at least five charged tracks, a total visible energy greater than 15\% of
the center-of-mass energy, and a primary vertex consistent with
the known collision point.
To suppress continuum background, we require the ratio $R_2=H_2/H_0$ of Fox-Wolfram 
moments\cite{foxwolfram}
 to be less than 0.4. 

\section{Analysis}
A full discussion of the partial reconstruction method as it has been applied
to the decay $\bar{B}^0 \rightarrow D^{*+} \ell^- \bar{\nu}_{\ell}(D^{*+} 
\rightarrow D^0 \pi^+)$ can be found in \cite{mssthesis}.
Here we give a brief description, with an emphasis on features that are
particularly relevant to this measurement.

\begin{figure}
%{\centerline{\hbox
{\includegraphics[height=7cm]{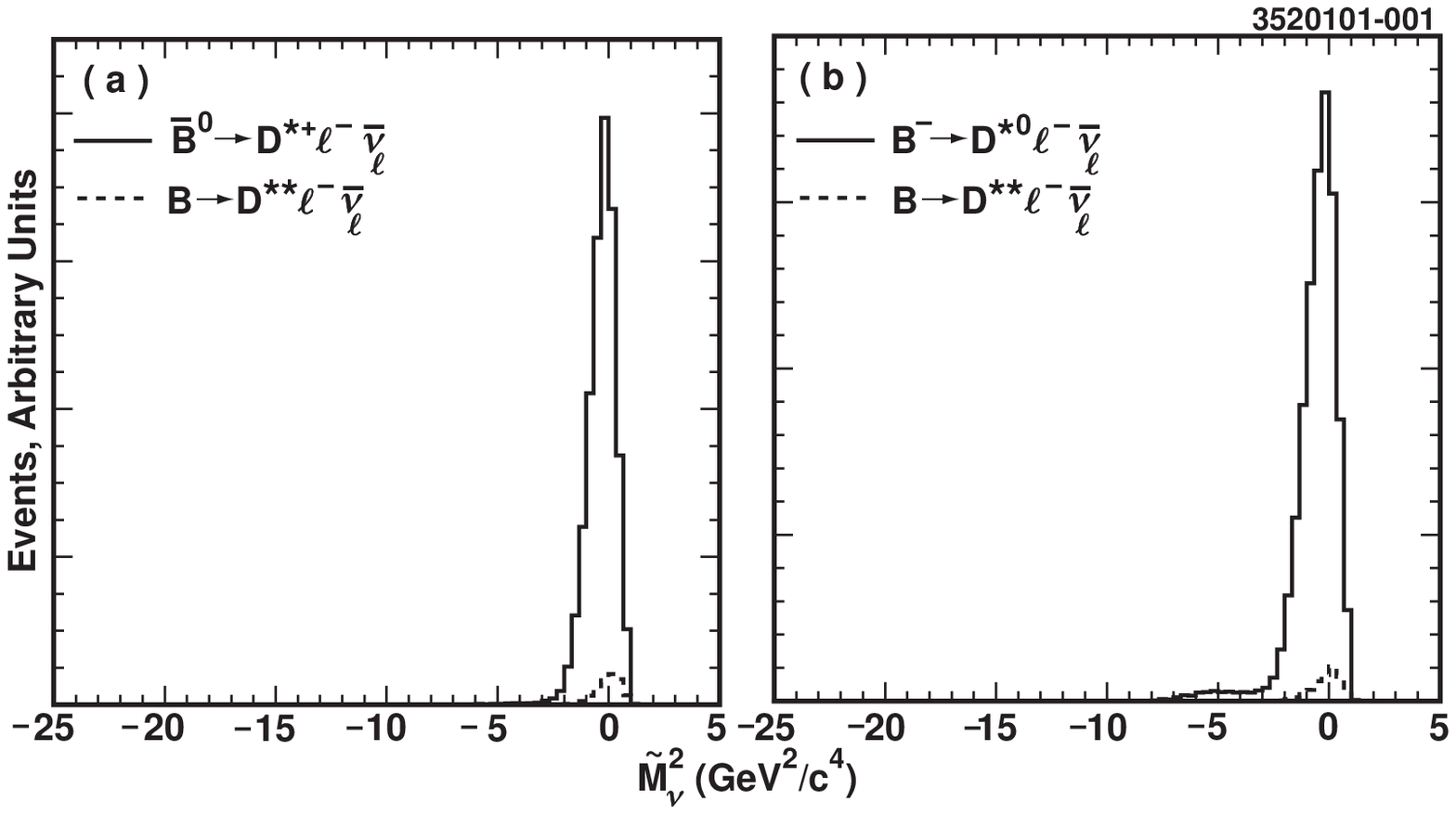}}
%{\psfig{figure=3520101-001.ps,height=7cm}}
%}}
\caption{\label{fig:mcrspipi0} The distribution in $\widetilde{\cal M}_\nu^2$ for signal
(solid histogram), obtained from MC simulation: 
(left) $\ell-\pi^+$ candidates from $\bar{B}^0 \rightarrow D^{*+} \ell^- \bar{\nu}_{\ell}$  \{$D^{*+} \rightarrow D^0 \pi^+$\};
(right) $\ell-\pi^0$ candidates from $B^- 
\rightarrow D^{*0} \ell^- \bar{\nu}_{\ell}$ \{$D^{*0} \rightarrow D^0 \pi^0$\} and $\bar{B}^0 
\rightarrow D^{*+} \ell^- \bar{\nu}_{\ell}$ \{$D^{*+} \rightarrow D^+ \pi^0$\}.
The dashed histogram shows the estimated relative contribution from decays
$\bar{B} \rightarrow D^{*} \pi \ell^- \bar{\nu}_{\ell}$ (correlated background).
}  

\end{figure}

A partially reconstructed decay $\bar B \rightarrow D^* \ell^- \bar{\nu}_{\ell}$
(inclusion of the charge conjugate mode is implied throughout
this report) 
consists of an identified lepton in combination with a
soft pion from the decay $D^*\rightarrow D\pi$.
The approximate four-momentum of the $D^*$, 
($\widetilde{E}_{{D^*}},\widetilde{\bf{p}}_{{D^*}}$),
is calculated by scaling the pion momentum:
\begin{eqnarray*}
E_{D^*} &\simeq& {E_{\pi}\over E^{CM}_{\pi}}M_{D^*} \equiv \widetilde{E}_{D^*},\ {\rm and}\\
{\bf{p}}_{D^*}&\simeq &
{\hat{\bf{p}}}_\pi\times{\sqrt{\widetilde{E}_{D^*}^2-M_{D^*}^2}}
\equiv\widetilde{\bf{p}}_{D^*},
\end{eqnarray*}
where $E_\pi$ is the the pion energy, $E^{CM}_{\pi} \approx 145$~MeV 
is the energy of the pion in the 
$D^{*}$ rest frame, and $M_{D^*}=2.01$~GeV/$c^2$ is the mass of 
the $D^{*}$. 
Using the 
approximation ${\bf p_B}\simeq 0$ we can calculate a ``squared missing mass'':
\begin{eqnarray}
\widetilde{\cal M}_\nu^2 \equiv (E_{\mbox{beam}}-\widetilde{E}_{{D^*}} - 
E_{\ell})^2-(\widetilde{\bf{p}}_{{D^*}} + {\bf{p}}_{\ell})^2,
\label{eqn:mm2}
\end{eqnarray}
where $E_{beam}$ is the beam energy and $E_{\ell}$ and ${\bf{p}}_{\ell}$ are the energy and momentum of the lepton.
Correctly identified signal candidates accumulate in the region
$\widetilde{\cal M}_\nu^2>-2.0 ($GeV$^2/c^4)$, 
as can be seen in Figure~\ref{fig:mcrspipi0}.
We define this to be the ``peak''  region.
The ``sideband'' region, defined as 
$-25.0<\widetilde{\cal M}_\nu^2<-4.0$~GeV$^2$, is used for estimating
backgrounds.

For $\bar{B}^0 \rightarrow D^{*+} \ell^- \bar{\nu}_{\ell}$, 
leptons are combined with charged pions.
Phase space limitations prohibit the 
decay chain $B^- \rightarrow D^{*0} \ell^- \bar{\nu}_{\ell}$, $D^{*0} \rightarrow 
D^+ \pi^- $, leaving only the $\bar{B}^0$ decay as a source of leptons 
which correlate with slow charged pions.  
In this decay, the lepton 
must have a charge opposite to that of the slow pion.  
This combination will 
be referred to as $\ell-\pi^+$.  

A lepton may also be combined with a slow neutral pion to reconstruct
decays $B\rightarrow D^*\ell\bar\nu$.
In this case both charged and neutral $B$'s contribute to the
signal, through the decays $B^- \rightarrow D^{*0} \ell^- \bar{\nu}_{\ell}(D^{*0} 
\rightarrow D^0 \pi^0)$ and $\bar{B}^0 \rightarrow 
D^{*+} \ell^- \bar{\nu}_{\ell}(D^{*+} \rightarrow D^+ \pi^0)$.
Accounting for the respective $D^*\rightarrow D\pi$ branching fractions,
the reconstructed signal is expected to include
$B^- \rightarrow D^{*0} \ell^- \bar{\nu}_{\ell}$ and  
$\bar{B}^0 \rightarrow D^{*+} \ell^- \bar{\nu}_{\ell}$ in a ratio of approximately 2:1. 
The value of $\widetilde{M}_{\nu}^2$ is calculated using the measured  momentum
of the $\pi^0\rightarrow\gamma\gamma$ candidate and the known $\pi^0$ mass.
Such combinations will be referred to as $\ell-\pi^0$.  

Lepton candidates are required to have momentum between 1.8~GeV/$c$ and 
2.5~GeV/$c$.  
Electrons are identified by using the ratio of the calorimeter 
energy to track momentum $(E/p)$ and specific ionization $(dE/dx)$ 
information.  
Muons are required to penetrate at least five
interaction lengths of absorber material.  
Electrons and muons must fall in the fiducial
region $|\cos \theta| < 0.707$, where $\theta$ is the polar angle of
the track's momentum vector with respect to the beam axis.

All charged pion candidates are required to have momentum between 90~MeV/$c$ and 
220~MeV/$c$ and to be consistent with originating at the interaction point.

Neutral pion candidates are obtained from energy deposits in the cesium iodide 
electromagnetic calorimeter which do not match the projection of any 
charged track and are consistent with being electromagnetic showers.  
Each shower must have energy greater than 50~MeV and
be in the good barrel region only ($\cos\theta_{\gamma} < 0.71$), where
$\theta_{\gamma}$ is the polar angle with respect to the beam axis.
The width of the $\gamma\gamma$ invariant mass of $\pi^0$'s 
depends on both the energy and angular resolution 
of the component photons and averages 5~MeV/$c^2$. 
In order to be able to evaluate the background from random $\gamma\gamma$ pairs (``fake $\pi^0$'s''),
we consider candidates within
a wide mass range, $0.085<M_{\gamma\gamma}<0.185$~GeV/$c^2$.

The $\ell-\pi^+$ and $\ell-\pi^0$ analyses differ
mainly in the detection and identification of the soft pion.
Because the efficiencies for $\pi^+$ and $\pi^0$ differ and 
are momentum-dependent, the overall reconstruction
efficiencies will also differ, and their calculation will depend on the decay 
model used to obtain them, due to differences in momentum distributions.
It is reasonable to assume, however, that the isospin-related modes
$B^- \rightarrow D^{*0} \ell^- \bar{\nu}_{\ell}$ and 
$\bar{B}^0 \rightarrow D^{*+} \ell^- \bar{\nu}_{\ell}$ have the same decay
dynamics, and, since the charged and neutral meson masses
are nearly equal for both $B$ and $D^*$, nearly identical kinematics.
The efficiencies to partially reconstruct the two modes should 
thus differ only due to the momentum-dependent pion efficiencies.
We therefore extract the signal yields and ratios in bins 
of 10~MeV/$c$ in the pion momentum
and are then able to make direct model-independent 
comparisons.

The signal is extracted by counting $\ell-\pi$ candidates which fall in the peak region
of $\widetilde{\cal M}_\nu^2$ (``peak candidates'') and evaluating and subtracting the backgrounds.
Backgrounds include contributions from continuum events, leptons
or $\pi^0$'s that are fake,
and real lepton--pion pairs in $B\bar B$ events that are not signal.
Some of the backgrounds are evaluated using Monte Carlo (MC) simulations,
where comparisons with candidates in the sideband region (``sideband candidates'') are used to obtain scaling factors.

The contribution from continuum events is evaluated 
by analyzing the off-resonance data and scaling to account 
for the differences in the integrated luminosity and center-of-mass energy.
The contribution from fake leptons in $B\bar B$ events  is estimated by counting continuum-subtracted candidates 
in which the
``lepton'' satisfies all criteria except lepton identification,
weighted by the fake probability, which is a function of
momentum \cite{fakprob}.
After subtracting continuum and fake-lepton backgrounds, the remainder 
consist of real leptons in combination with soft pion candidates in $B\bar B$ events.
For $\ell-\pi^+$ the numbers are obtained through simple counting.
In the case of $\ell-\pi^0$ there is additional background from fake $\pi^0$'s,
formed by random photon combinations.
The yields are therefore extracted by fitting the ${\gamma\gamma}$ invariant mass distribution.  
We first discuss the remaining backgrounds for $\ell-\pi^+$ and then
return to the treatments of $\ell-\pi^0$.

For the $\ell-\pi^+$ case the numbers after continuum and fake lepton
subtractions are those of peak and sideband candidates 
consisting of a real lepton  in combination 
with a real soft pion from a $B\bar B$ event.
The remaining backgrounds are of two general types.
Candidates formed from the lepton and soft pion from the $D^*$ in
decays of the type
$\bar{B} \rightarrow D^{*} \pi \ell^- \bar{\nu}_{\ell}(D^{*}\rightarrow D^0\pi)$ will
accumulate near $\widetilde{\cal M}_\nu^2=0$ and are
difficult to distinguish from signal.
We define this type as ``correlated background.''
All other combinations, where the pion is not the daughter of a $D^*$ from
the same $B$ meson as the lepton, are defined as ``uncorrelated background.''

We use the CLEO MC simulation, the decay model of Goity and Roberts\cite{goityroberts}, and measured branching fractions 
to estimate the contribution of correlated background. 
The $D^*\pi$ combinations include
three $D^{**}$ resonances, as well as non-resonant production.
The total rate is set to match the ALEPH measurement\cite{lepd2s},
$ {\cal B}(b\rightarrow \bar B)\times{\cal B}(\bar B\rightarrow D^{*+}\pi^-\ell^-\nu X)=
(4.73\pm 0.77\pm 0.55)\times 10^{-3}$.
By isospin symmetry, we assume that $B\rightarrow D^{*+}\pi^0\ell^+\nu X$
contributes additionally at half of this rate, and that the rates to $D^{*0}$ are
equal to the corresponding rates for $D^{*+}$.
We then use the ALEPH value ${\cal B}(b\rightarrow \bar B)=(37.8\pm 2.2)\%$ to obtain
${\cal B}(B\rightarrow D^{*-/0}\pi\ell^+\nu X)=
(1.88\pm 0.39)\%$, where the statistical and systematic errors
have been added in quadrature.
The resulting numbers of candidates are scaled to the integrated
luminosity of the data and subtracted directly from the peak
and sideband samples.
The $D^{**}$ contributions to the peak are small, due to the  high lepton momentum requirement 
imposed in this analysis.
We find that this background comprises 
$(2.8\pm 0.2)$\% of the net signal.
The contributions to the sideband are much smaller.

\begin{figure}
{\centerline{
%{\psfig{figure=3520101-003.ps,height=10cm}}
 \includegraphics[height=10cm]{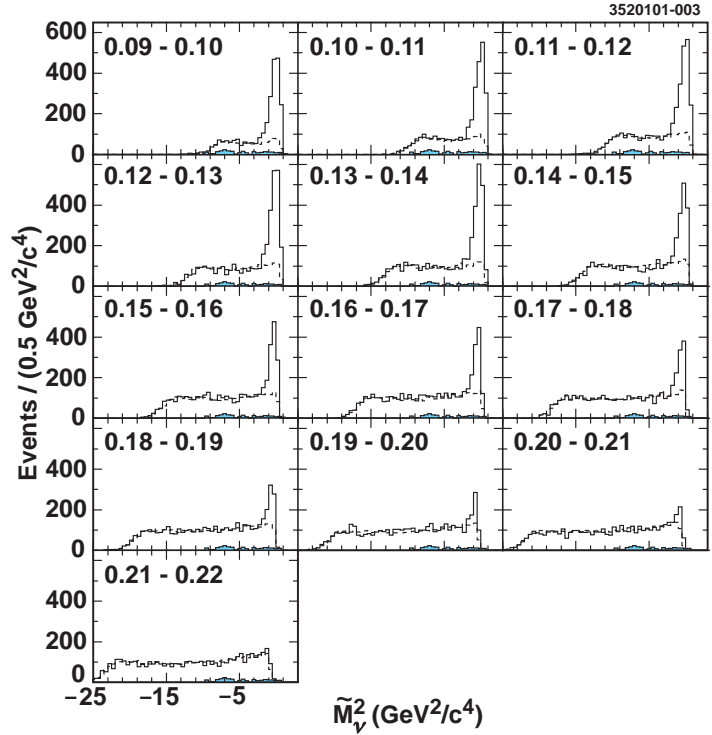}
}}
\caption{\label{fig:mompi} Distributions in $\widetilde{\cal M}_\nu^2$ for the decay $\bar{B}^0 
\rightarrow D^{*+} \ell^- \bar{\nu}_{\ell}$, for 10~MeV/c bins of pion momentum.
The plot shows right sign data, with continuum and fake lepton contributions subtracted
(solid histograms), 
uncorrelated background (dashed), and correlated backround (black filled)
estimated by  Monte Carlo.}  
\end{figure}

The uncorrelated background is estimated via MC simulation. 
We reconstruct $\ell-\pi^+$ candidates in 17.5~million 
generic MC 
$B\bar{B}$ events, excluding signal and correlated background.
To obtain a data/MC scaling factor, ratios of sideband candidates in 
data and MC are found in all momentum bins and the set of 
ratios is fitted to a single value.
The numbers of peak candidates from MC are multiplied 
by this factor to obtain the contributions of uncorrelated background.
Figure~\ref{fig:mompi} displays the $\widetilde{\cal M}_\nu^2$ distributions
in each pion momentum bin of data after continuum
and fake subtractions, correlated background (MC), and uncorrelated
background (MC, scaled).
The agreement between MC and data of the uncorrelated background
in the sideband region is excellent.
Peak candidates in excess of the evaluated backgrounds
comprise our signal.

\begin{figure}
%{\centerline{\hbox
%{\psfig{figure=3520101-004.ps,height=8cm}}
\includegraphics[height=9cm]{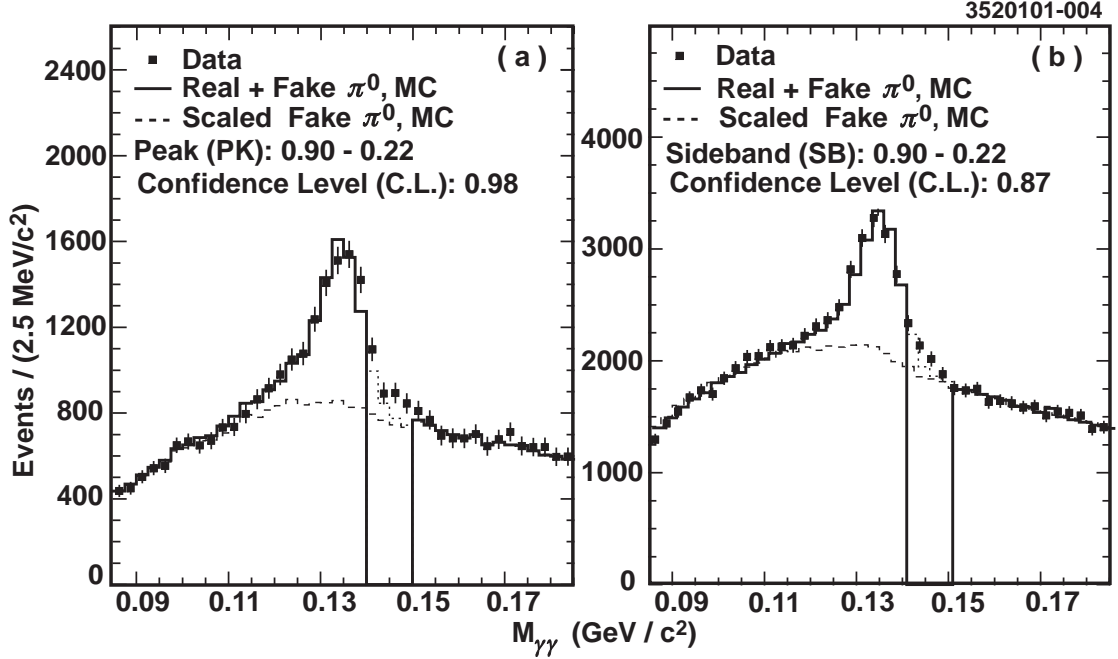}
%}}
\caption{\label{fig:pi0fakeunbin} Fit of $M_{\gamma\gamma}$ distribution comprising
$\ell-\pi^0$ peak (left) and sideband (right) tags.  
The region 0.140-0.150~GeV/c$^2$ is omitted from the fit in this
and all other fits of $M_{\gamma\gamma}$ described in this report.}
\end{figure}

\begin{figure}
%{\centerline{\hbox
%{\psfig{figure=3520101-005.ps,height=8cm}}
 \includegraphics[height=9cm]{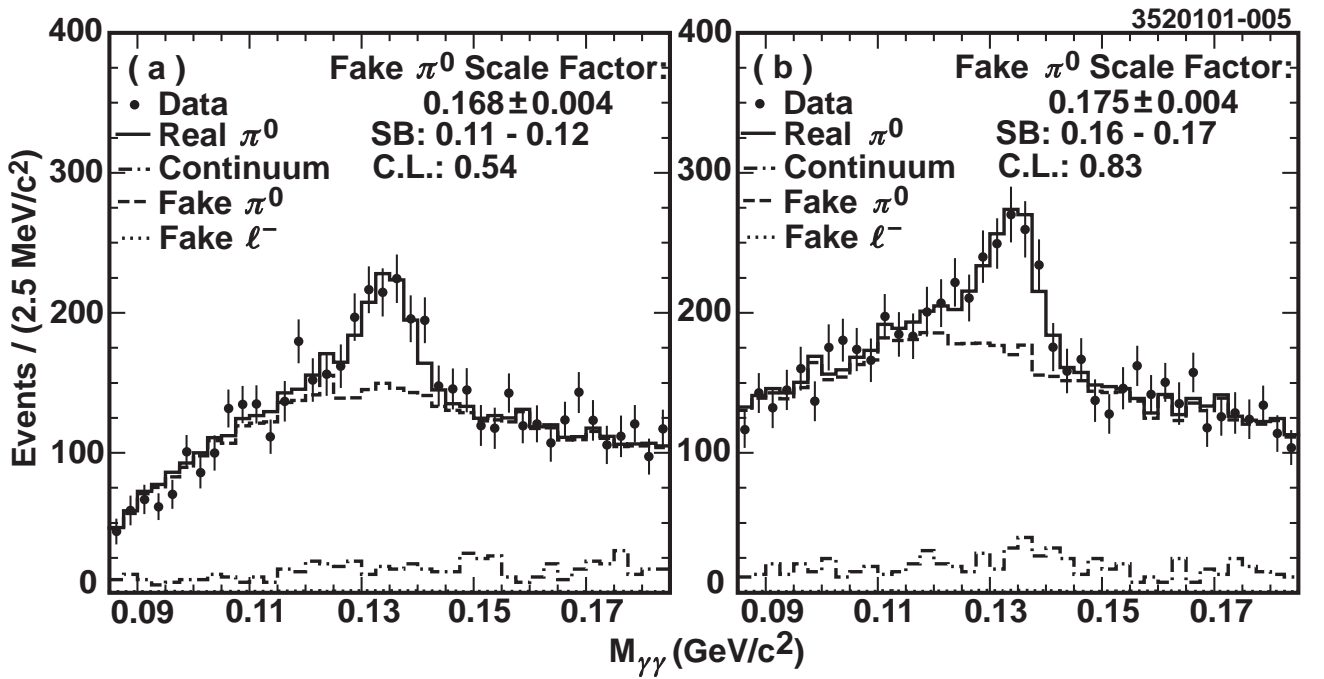}
%}}
\caption{\label{fig:pi0fake1} Fits of $M_{\gamma\gamma}$ to MC-generated real 
and fake $\pi^0$ distributions
for $\ell-\pi^0$ sideband candidates, 0.11-0.12~GeV/c$^2$ (left) and 0.16-0.17~GeV/c$^2$ (right).
The fake lepton contributions are not visible in this
plot.
This fit was used to determine the data/MC scaling factor
for fake $\pi^0$'s.
}
\end{figure}

\begin{figure}
%{\psfig{figure=3520101-006.ps,height=8cm}}
\includegraphics[height=9cm]{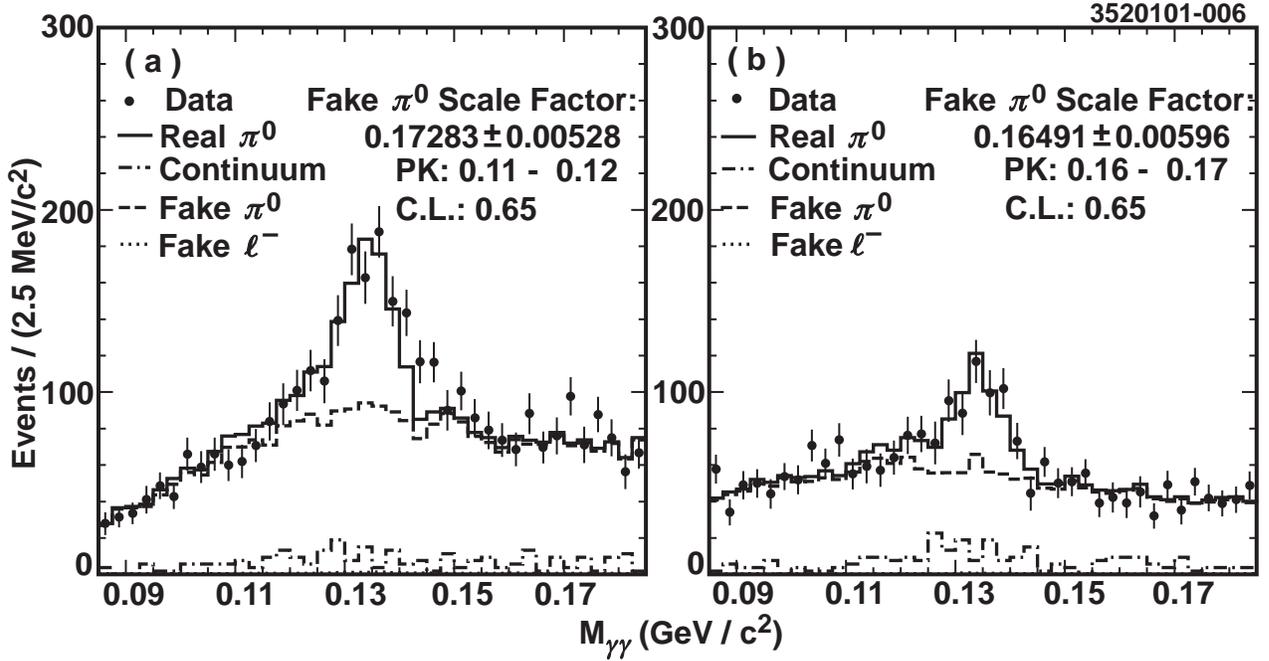}%
\caption{ \label{fig:pi0fake2} Plots as described in Figure~\ref{fig:pi0fake1}, for
peak candidates.
}
\end{figure}

\begin{figure}
%{\psfig{figure=3520101-007.ps,height=8cm}}
\includegraphics[height=8cm]{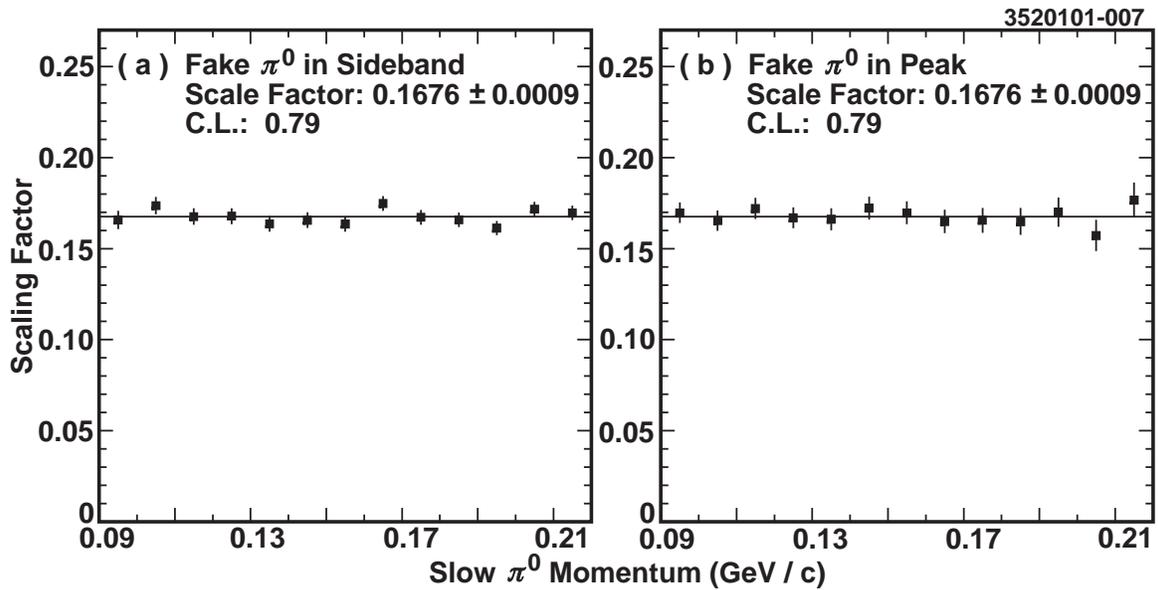}%
\caption{\label{fig:pi0fake3} Fit of fake $\pi^0$ scaling factors to a constant.  
Scaling factors are
plotted as a function of momentum for sideband (left) and peak (right) candidates
and fitted simultaneously to a single constant.
}
\end{figure}

The procedure applied to $\ell-\pi^0$ is similar in concept,
but instead of simple counting
we fit the $\gamma\gamma$ invariant mass distributions.
After the direct subtractions of continuum and
fake leptons, the remaining distributions contain in addition to the backgrounds
discussed for the $\ell-\pi^+$ (correlated and uncorrelated background)
the contribution from $B\bar B$ events where the lepton is
real and the pion is fake.
The correlated background is estimated and subtracted first.
In the case of $\ell-\pi^0$ there are contributions from modes containing
both charged and neutral $D^*$.
As with $\ell-\pi^+$, the contributions are calculated based on the
branching fraction measured at LEP and MC simulation.
We find that the correlated background comprises $(2.5\pm 0.9)$\% of the net signal
in the peak region.
The contribution in the sideband region, which is much smaller, is also calculated and
subtracted.
We then determine the contribution of fake $\pi^0$.
The resulting $\gamma\gamma$ invariant mass distributions are fitted
 to a sum of real and fake $\pi^0$ distributions generated
via MC simulation. 
In the fit we exclude the region 
$0.14<M_{\gamma\gamma}c^2<0.15$~GeV because the simulated $\pi^0$
signal in this region does not show good agreement 
with data.
Although this disagreement is not apparent in the individual pion momentum
bins, due to insufficient statistics, it is revealed when we 
fit to a distribution that is summed
over all bins (Figure~\ref{fig:pi0fakeunbin}). 
We find that the fake $\pi^0$ background is well simulated by MC in
that the fits all have good confidence and that the data/MC scaling factors 
are consistent with being a single constant over all $\pi^0$ momenta
for both sideband and peak distributions.
Figures~\ref{fig:pi0fake1} and~\ref{fig:pi0fake2} show 
examples of fit results.
Figure~\ref{fig:pi0fake3} shows results from a fit of all of the
scaling factors to a single constant.
That fit gives a confidence level of 80\% and a scaling factor 
of $0.1676 \pm 0.0009$, very close to the 
ratio of the numbers of $B\bar B$ events in data and MC
($0.1667 \pm 0.0001$).
We use this fitted value to calculate the 
amount of fake $\pi^0$ in each distribution considered.

\begin{figure}
%{\psfig{figure=3520101-008.ps,height=8cm}}
 \includegraphics[height=8cm]{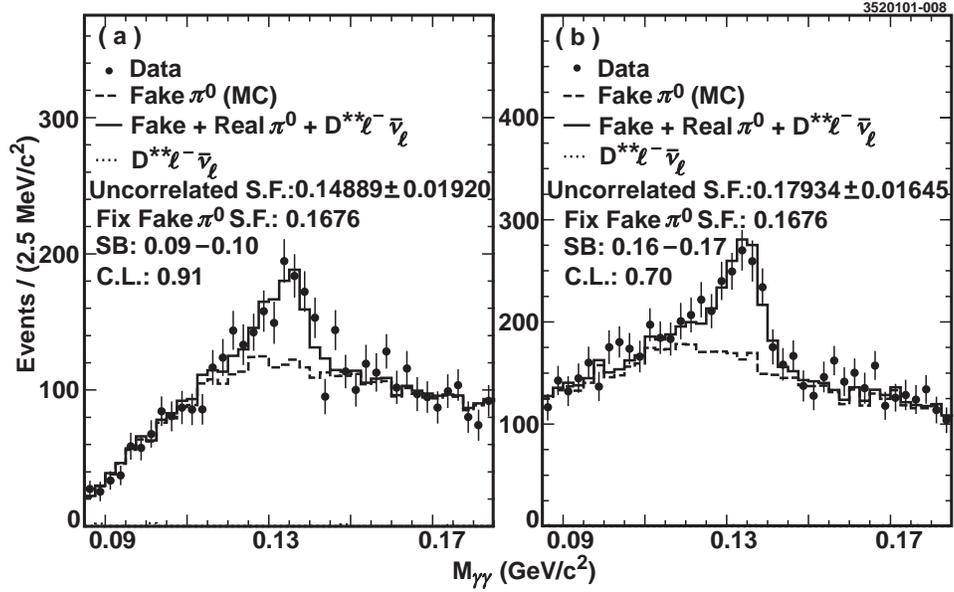}%
\caption{\label{fig:sbbgfit} Fits of $M_{\gamma\gamma}$ for sideband candidates,
to determine data/MC normalization of uncorrelated background.
The fake normalization is fixed.  The correlated background
from $D^{**}\ell\nu$ is not visible in this plot.
Shown are plots for two bins of $\pi^0$ momentum.}
\end{figure}

\begin{figure}
%{\psfig{figure=3520101-009.ps,height=9cm}}
 \includegraphics[height=9cm]{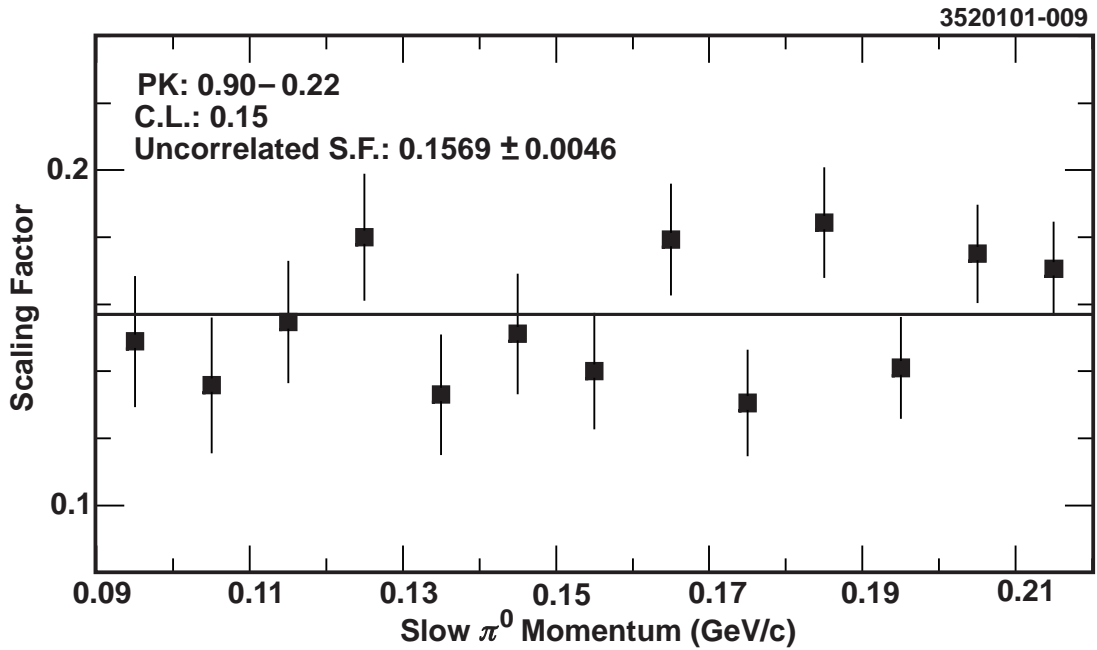}%
\caption{\label{fig:bgsffit} Fit of data/MC scaling factors for uncorrelated background.
Scaling factors obtained by fitting in momentum bins are 
fitted to a single constant.}
\end{figure}

The sideband candidates  remaining after the above background subtractions
consist of uncorrelated background.
For each momentum bin, the associated $M_{\gamma\gamma}$ distribution is
fitted to the corresponding MC simulation to obtain a data/MC scaling factor.
Figure~\ref{fig:sbbgfit} shows two plots with typical fit results.
A fit of the resulting scaling factors
(Figure~\ref{fig:bgsffit}) shows good consistency with the hypothesis of a single constant scaling factor.
The fitted value of $0.157\pm 0.005$ is used as the overall scaling factor to estimate the
uncorrelated background
in the peak region.

\begin{figure}
%{\psfig{figure=3520101-010.ps,height=8cm}}
 \includegraphics[height=9cm]{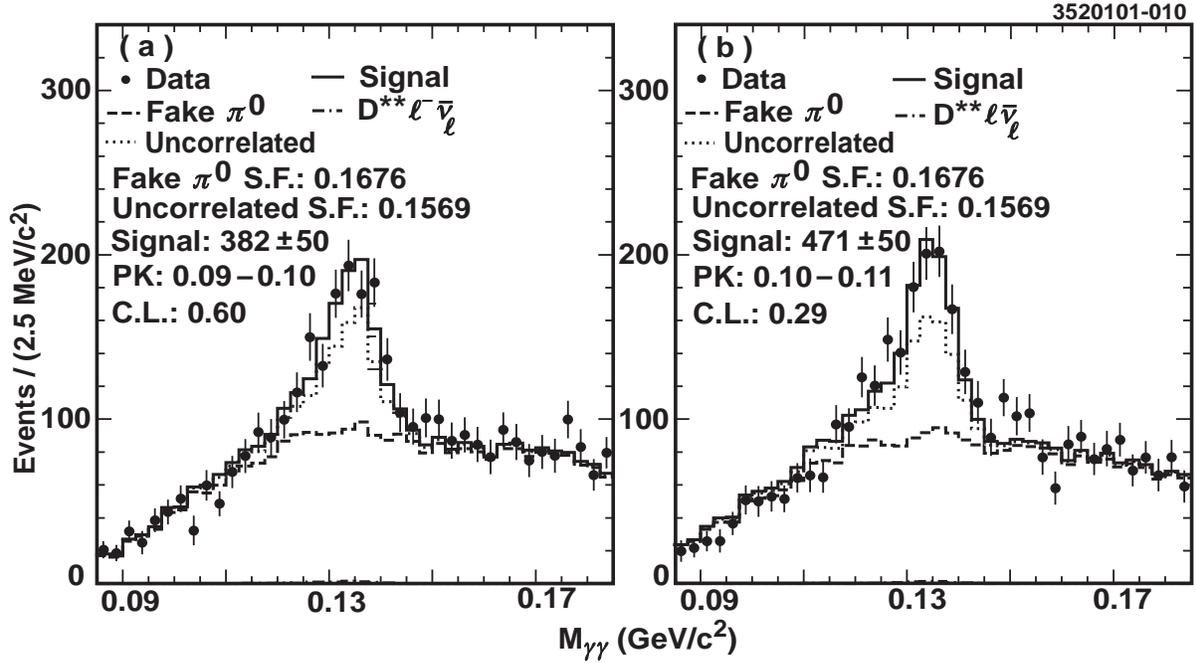}%
\caption{\label{fig:ggfit2sg} Fits of $M_{\gamma\gamma}$ for peak candidates,
to determine signal yields.
The fake, correlated, and uncorrelated background normalizations are fixed.
Shown are plots for two bins of $\pi^0$ momentum.}
\end{figure}

We fit the remaining $M_{\gamma\gamma}$ distribution
of peak candidates in each momentum bin to obtain the signal yield.
Two examples are shown in Figure~\ref{fig:ggfit2sg}. 

\subsection{Signal Yields}

\begin{figure}
%{\psfig{figure=3520101-011.ps,height=8cm}}
 \includegraphics[height=8cm]{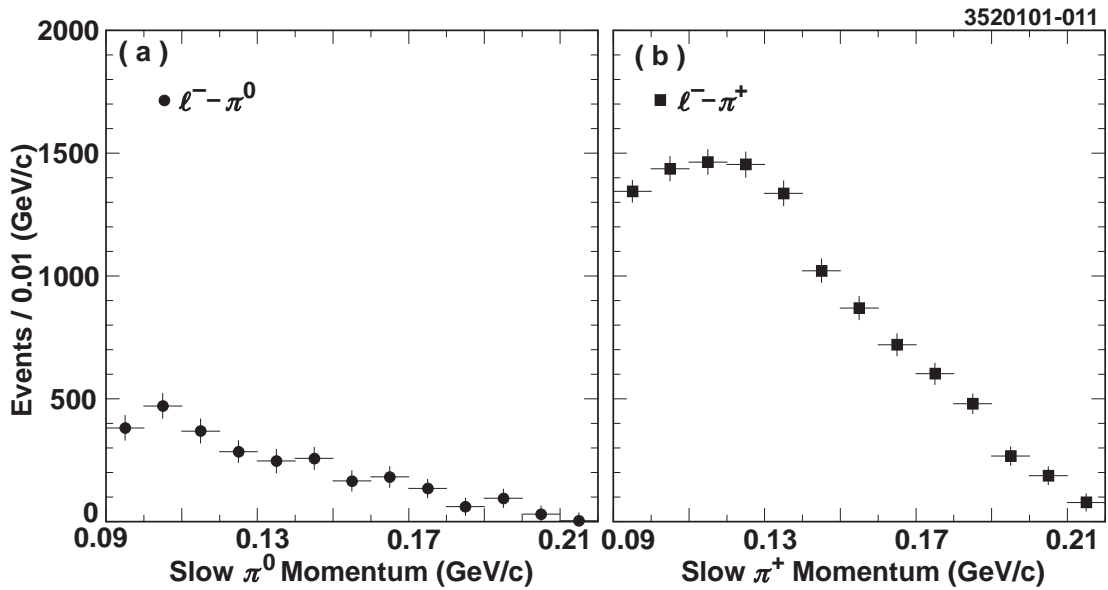}%
\caption{\label{fig:siglpibin} Yield in data of partially reconstructed signal
in $\ell-\pi^0$(left) and
$\ell-\pi^+$ (right) as a 
function of $\pi$ momentum. 
Distributions are not corrected for 
reconstruction efficiency.}
\end{figure}

After all background subtractions listed above, 
a total of $11,262 \pm 164$ $\ell-\pi^+$ and
$2,686\pm 142$ $\ell-\pi^0$  remain as signal. 
Figure~\ref{fig:siglpibin} shows the yields as a function 
of pion momentum.

\begin{figure}
%{\psfig{figure=3520101-012.ps,width=10cm}}
 \includegraphics[height=10cm]{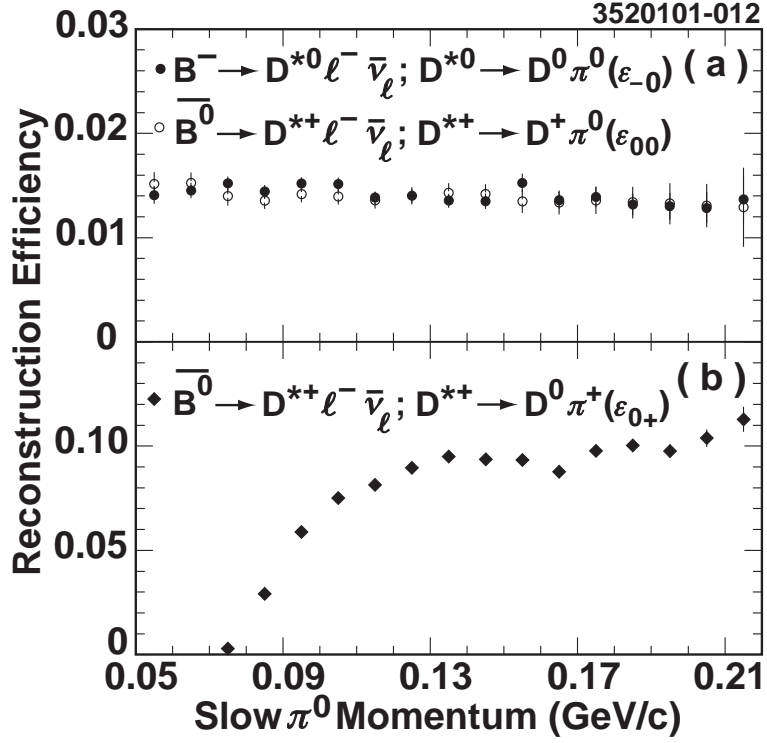}%
\caption{\label{fig:lpi0effs} Efficiencies for reconstruction of decays via partial reconstruction,
as a function of $\pi$ momentum:
$\ell-\pi^0$ (top) and $\ell-\pi^+$ (bottom) analysis. }  
\end{figure}

To obtain reconstruction efficiencies, 
we generated events containing
$B\rightarrow D^*\ell\bar\nu(D^*\rightarrow D\pi)$ using the model of Scora and Isgur\cite{isgw2} (ISGW2).
The events were passed through a full GEANT-based detector simulation
and offline analysis.  
We perform the analysis on tagged signal candidates.
Figure~\ref{fig:lpi0effs} shows the resulting efficiency as a function of pion momentum 
for both $\ell-\pi^+$ and $\ell-\pi^0$ analyses.
The efficiencies for $\ell-\pi^0$ reconstruction of
$\bar{B}^0 \rightarrow D^{*+} \ell^- \bar{\nu}_{\ell}$
and 
$\bar{B}^- \rightarrow D^{*0} \ell^- \bar{\nu}_{\ell}$
are expected to be equal, and since the MC result is consistent
with this expectation, it is assumed to be the case.

\section{Relationship of measurement to decay rates}

The numbers of reconstructed $\ell-\pi^+$ and $\ell-\pi^0$ signal, 
$N_{+}$ and $N_0$,
have the following relationships to the branching fractions:
\begin{eqnarray}
N_+ & = & \nonumber \\
 & & 2 \times 2 \times N_{B\bar{B}}f_{00} \times 
                          \epsilon_{0+} \times 
           {\cal B}(\bar{B}^0 \rightarrow D^{*+} \ell^- \bar{\nu}_{\ell}) \times 
                    {\cal B}(D^{*+} \rightarrow D^0 \pi^+),\\ {\rm and} \label{eq:nspi}\\
N_0 & = & \nonumber \\
& & 2 \times 2 \times N_{B\bar{B}}f_{+-} \times 
                    \epsilon_{-0} \times 
               {\cal B}(B^- \rightarrow D^{*0} \ell^- \bar{\nu}_{\ell})\times 
                          {\cal B}(D^{*0} \rightarrow D^0 \pi^0)\nonumber \\ 
                    & +  & 2 \times 2 \times N_{B\bar{B}}f_{00} \times 
                          \epsilon_{00} \times 
                {\cal B}(\bar{B}^0 \rightarrow D^{*+} \ell^- \bar{\nu}_{\ell}) 
\times  
                    {\cal B}(D^{*+} \rightarrow D^+ \pi^0), \label{eq:nspi0}   
\end{eqnarray} 
where $f_{00}$ $(f_{+-})$ is the fraction of neutral (charged) B mesons 
in $\Upsilon(4S)$ events, and $\epsilon_{0+}$,  $\epsilon_{-0}$,  and $\epsilon_{00}$
are the reconstruction efficiencies for the respective $B\rightarrow D^*\ell\nu
(D^*\rightarrow D\pi)$ modes.
Two factors of 2 enter in these expressions because each $B\bar{B}$ event 
contains two B mesons, and we add the signals for electrons and muons.
Solving for the branching fractions,
\begin{eqnarray}
{\cal B}(\bar{B}^0 \rightarrow D^{*+} \ell^- \bar{\nu}_{\ell}) &=& 
\frac{N_+}{4 N_{B\bar{B}}f_{00} 
\epsilon_{0+} {\cal B}(D^{*+} \rightarrow D^0 \pi^+)} 
\label{eq:3}\\
{\cal B}(B^- \rightarrow D^{*0} \ell^- \bar{\nu}_{\ell}) &=& 
\frac{N_0 - 4 N_{B\bar B}f_{00} 
\epsilon_{00} {\cal B}(\bar{B}^0 \rightarrow 
D^{*+} \ell^- \bar{\nu}_{\ell}) {\cal B}(D^{*+} \rightarrow D^+ \pi^0)}
{4 N_{B\bar{B}}f_{+-}\epsilon_{-0} {\cal B}(D^{*0} 
\rightarrow D^0 \pi^0)}\nonumber \\ 
&=&
\frac{N_0 - 
N_+ {\epsilon_{00} {\cal B}(D^{*+} \rightarrow D^+ \pi^0)\over
\epsilon_{0+} {\cal B}(D^{*+} \rightarrow D^0 \pi^+)}}
{4 N_{B\bar{B}}f_{+-}\epsilon_{-0} {\cal B}(D^{*0} 
\rightarrow D^0 \pi^0)}\label{eq:4}
\end{eqnarray}
Dividing equation~(\ref{eq:4}) by equation~(\ref{eq:3}), assuming that 
$\epsilon_{-0}=\epsilon_{00}$, and defining 
$n\equiv{N_+\epsilon_{00}\over N_0\epsilon_{0+}}$, we get
\begin{eqnarray}
{{\cal B}(B^- \rightarrow D^{*0} \ell^- \bar{\nu}_{\ell})\over 
{\cal B}(\bar{B}^0 \rightarrow D^{*+} \ell^- \bar{\nu}_{\ell})}
&=&
{f_{00}[{\cal B}(D^{*+} \rightarrow D^0 \pi^+)
-n{\cal B}(D^{*+} \rightarrow D^+ \pi^0)]\over 
nf_{+-}{\cal B}(D^{*0} \rightarrow D^0 \pi^0)},\ {\rm and} \\
\alpha\equiv {f_{+-}\over f_{00}}{{\cal B}(B^- \rightarrow D^{*0} \ell^- \bar{\nu}_{\ell})\over 
{\cal B}(\bar{B}^0 \rightarrow D^{*+} \ell^- \bar{\nu}_{\ell})}
&\equiv&
{{\cal B}(D^{*+} \rightarrow D^0 \pi^+)
-n{\cal B}(D^{*+} \rightarrow D^+ \pi^0)\over 
n{\cal B}(D^{*0} \rightarrow D^0 \pi^0)}.
\end{eqnarray}

\begin{figure}
%{\psfig{figure=3520101-013.ps,height=8cm}}
 \includegraphics[height=8cm]{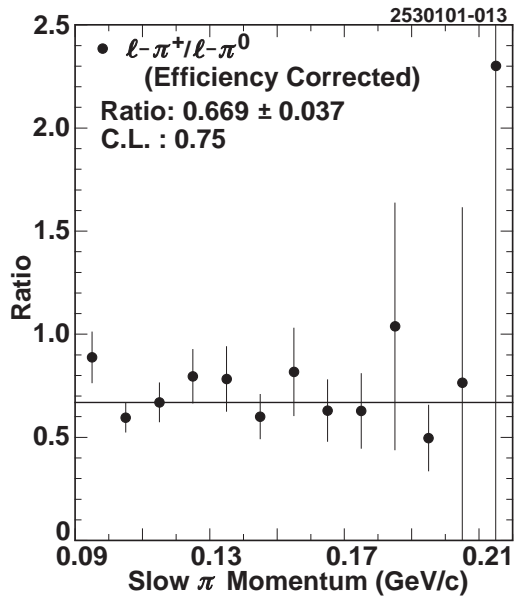}%
\caption{\label{fig:n} Efficiency-corrected ratio of $\ell-\pi^+$ to $\ell-\pi^0$ candidates
in bins of $\pi$ momentum.  The values are fitted to a constant.}
\end{figure}

Although it is implied in the above equations that the ratio $n$ is
a ratio of total rates, it is equal to the ratio of rates over any
given restricted kinematic region, as long as the kinematics of
the decays are the same.
We thus take the ratio of the $\ell-\pi^+$ to $\ell-\pi^0$ signals as a 
function of pion momentum, correct for the reconstruction 
efficiencies in each bin to obtain $n$, and fit to a constant.
The fit yields an overall value  
$n=0.669\pm 0.037$ (Figure~\ref{fig:n}).
Using this value and the $D^*$ branching fractions shown in
Table~\ref{tab:dick}, we obtain 
$\alpha=1.136\pm 0.090$, where the error is statistical only.

\section{Systematic Errors}

The systematic uncertainty on $\alpha$ is due to the uncertainty in determining
the ratio $n$ and uncertainties in the $D^*$ branching fractions.
The reconstructions $\ell-\pi^0$ and $\ell-\pi^+$ have
many features in common and therefore many shared systematic errors
which cancel at least partially in taking the ratio.
The principal difference between the two is in the
reconstruction of the pions.

We first discuss the uncertainties from lepton detection/identification
and background subtractions
that are common to both reconstructions.
The uncertainty on the continuum subtraction is obtained by
changing the scaling factor up and down by 3\%.
The uncertainty on the lepton fake probability
is 30\%.
The systematic uncertainty on the lepton identification efficiency 
has been determined to be $2.5\%$\cite{mssthesis}. 
The systematic uncertainties from each of these sources is estimated
by varying the affected quantity and observing the shift of the result.
These errors cancel at least partially in taking the ratio $n$.

\begin{figure}
%{\psfig{figure=3520101-002.ps,height=9cm}}
 \includegraphics[height=9cm]{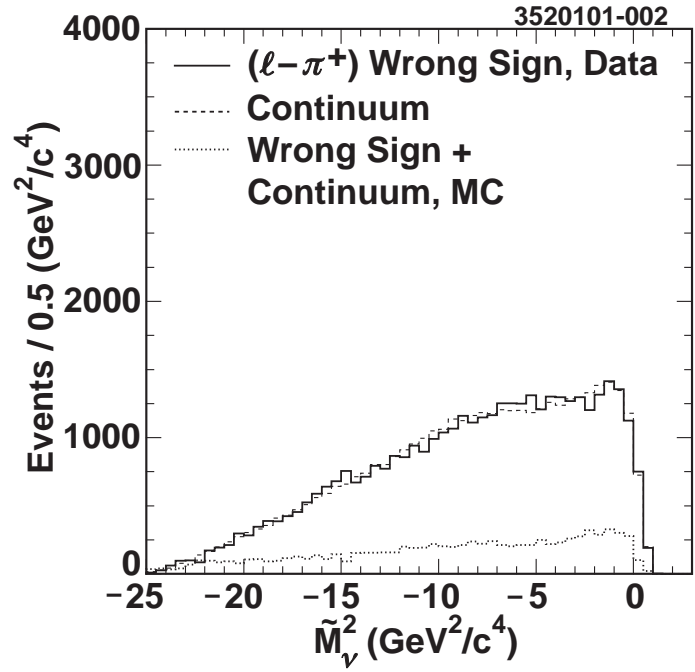}%
\caption{\label{fig:siglpi2} The distribution in $\widetilde{\cal M}_\nu^2$ for wrong-sign
$\ell-\pi^+$ candidates, data on resonance (solid histogram), 
data off resonance, scaled  (dotted), and MC normalized to
data in the sideband region (dashed--dotted).}
\end{figure}

The estimation of uncorrelated background depends on accurate
modeling of the peak/sideband ratio by MC.
The overall $\widetilde{\cal M}_\nu^2$ distribution 
of the uncorrelated background
is dominated by phase space, i.e., a hard lepton and soft pion
distributed isotropically will produce a distribution similar to
that shown.
However, because this analysis has high statistical precision, we 
are somewhat sensitive to
the finer details of the Monte Carlo $B$ decay generator.
A simple overall test of its reliability is the counting of
wrong sign ($\ell^\pm-\pi^\pm$) candidates, where 
no signal is expected.  
An analysis identical to that performed with right sign
$\ell-\pi^+$ yields a signal of $-91 \pm 117$, consistent
with zero.  
The wrong sign distribution of $\widetilde{\cal M}_\nu^2$ in data and 
MC simulation are shown in Figure~\ref{fig:siglpi2}.
We take the absolute sum of the ``signal'' and statistical error, 
$\left|{-91}\right| + 117 = 208$, 
as an estimate of the systematic uncertainty due to modeling.
This is 3.3\% of the number of wrong sign peak candidates after
continuum and fake subtractions, so we take 3.3\% as the fractional
uncertainty on the uncorrelated background in the right sign.
This translates to 1.9\% of the net signal in  $\ell-\pi^+$.
For the $\ell-\pi^0$, the corresponding uncertainty is 2.0\% of the 
net signal.
Although these two uncertainties are believed to be largely correlated,
and should cancel at least partially in taking the ratio, we 
conservatively take the error on $n$ to be equal to the larger one, 2.0\%.

To estimate the uncertainty due to our lack of knowledge of the $B$
semileptonic decays to $D^{*}\pi$, we vary both the total branching fraction
and the mix of resonant and non-resonant contributions to $D^{*}\pi$.
The branching fraction, which we took to be $(1.88\pm 0.39)$\% for
each $D^*$ charge, averaged over $B$ charges as explained earlier, is 
varied up and down by the amount of the error, holding fixed
the relative contributions of the different modes.
We recorded excursions of 0.5\% and 0.3\% for $\ell-\pi^0$ and
$\ell-\pi^+$, respectively.
We also allow each mode in turn to saturate the rate and take the
maximum excursion of the result as a systematic uncertainty.
We thus obtain errors of 0.8\%, 1.4\%, and 1.0\%, for
$\ell-\pi^0$, $\ell-\pi^+$, and $n$.

The systematic uncertainty on tracking efficiency for slow charged pions has 
been measured in a previous CLEO analysis\cite{barish} to be 5\%.  
The uncertainty on the efficiency ratio of slow neutral 
pions to slow charged pions was determined to be 7\%. 
As the absolute efficiency for neutral pion reconstruction is determined
by the charged pion efficiency and the ratio, the quadratic sum of their
uncertainties gives the systematic uncertainty of $8.6\%$\cite{barish}
on the neutral pion efficiency.  

To estimate the systematic error on the 
evaluation of fake $\pi^0$'s in $\ell-\pi^0$, we vary
the fake scaling factor up and down by the amount of its statistical error and
repeat the analysis. 
We take the maximum excursion of 1.6\% to be the systematic uncertainty from this source.
We also repeat the analysis without excluding the $\gamma\gamma$ mass
region $0.014-0.015$~GeV/c$^2$ and find that this shifts the
result by 0.7\%.
We add the two numbers quadratically to get a systematic error on
the $\pi^0$ fit of 1.7\%.

We find the total systematic errors to be $9.4\%$ 
for $\ell-\pi^0$, ${6.1}\%$ for 
$\ell-\pi^+$, and 7.6\% for $n$.
This results in an uncertainty of 10.9\% on $\alpha$.
The additional uncertainty from the $D^*$ branching fractions
is 6.3\%. 
All the systematic uncertainties are summarized in Table~\ref{tab:systerr}.

%----------------------------

\section{Summary and Conclusions}

By measuring the ratio of partially reconstructed $B\rightarrow D^*\ell\nu$
decays in the $\ell-\pi^+$ and $\ell-\pi^0$ channels as a function of
pion momentum, we obtain a measurement of the ratio
\begin{eqnarray*}
\frac{f_{+-}}{f_{00}} \frac{\tau_{+}}{\tau_{0}}=
1.136 \pm 0.090 \pm 0.143. 
\end{eqnarray*}
This result is in good agreement with published CLEO values 
(Table~\ref{tab:fpfm}).

Using the ratio of $B^+$ and $B^0$ lifetimes from a recent world 
average\cite{pdg}, 
\begin{eqnarray*}
 \frac{\tau_{+}}{\tau_{0}} = 1.074 \pm 0.028,
\end{eqnarray*}
we obtain the ratio of the charged and neutral $B$ meson production at 
the $\Upsilon(4S)$ resonance:
\begin{eqnarray*}
 \frac{f_{+-}}{f_{00}} = 1.058 \pm 0.084\pm 0.115 \pm 0.067 \pm 0.028
\end{eqnarray*}
where the first error is statistical, the second is systematic on n, 
the third is due to the uncertainty on the $D^*$ branching fractions,
and the fourth is due to the uncertainty on the $B$ lifetime ratio. 
We add the systematic errors quadratically to obtain
\begin{eqnarray*}
 \frac{f_{+-}}{f_{00}} = 1.058 \pm 0.084\pm 0.136.
\end{eqnarray*}

%----------------------------------------------

\begin{table}
\centering\caption{Branching fractions of $D^* \rightarrow D \pi$}
\begin{tabular}{ccc}
MODE     & Branching Fraction & reference \\ \hline
$D^{*+} \rightarrow D^0 \pi^+$   &  $(67.6 \pm 0.8)\%$ & \cite{Dstbr1} \\ 
$D^{*+} \rightarrow D^+ \pi^0$   &  $(30.7 \pm 0.7)\%$ & \cite{Dstbr1}  \\ 
$D^{*0} \rightarrow D^0 \pi^0$   &  $(61.9 \pm 2.9)\%$ & \cite{pdg} 
\end{tabular}
\label{tab:dick}
\end{table}

\begin{table}
\centering\caption{Summary of systematic errors (\%). }
\begin{tabular}{c c c c c} 
Source & $\ell-\pi^0$ & $\ell-\pi^+$ & $n$ & $\alpha$ \\ \hline
Continuum subtraction        & $0.2$  &  $0.2$  & 0.2 & 0.3  \\
Fake leptons                 & $0.5 $  &  $0.5 $ & $0.5 $ & $0.7 $ \\
Uncorrelated background      & $2.0 $  &  $1.9 $  & 2.0  & 2.9 \\ 
$D^{**}$ background           &$0.8 $  &  $1.4 $  & 1.0  & 1.4  \\
Fake $\pi^0$ subtraction      & $1.7 $   &  --  & 1.7  & 2.4 \\
$\pi$ efficiency         & $8.6 $  &  $5.0 $ & 7.0   & 10.0 \\
Lepton ID efficiency         & $2.5 $  &  $2.5 $  & -- & -- \\
$D^{*}$ Branching Fraction     & $-$  &  $-$  & $-$ & 6.3  \\ \hline
Total                        & $9.4 $  &  $6.1 $ & 7.6  & 12.6 
\end{tabular}
\label{tab:systerr}
\end{table} 

\begin{table}
\centering\caption{Other CLEO measurements of $\frac{f_{+-}}{f_{00}} 
\frac{\tau_{+}}{\tau_{0}}$}
\begin{tabular}{l c}
Mode       & $\frac{f_{+-}}{f_{00}} \frac{\tau_{+}}{\tau_{0}}$ \\ \hline
$B\rightarrow D^*\ell\bar\nu $\cite{barish}  & $1.14 \pm 0.14 \pm 0.13$ \\
$B \rightarrow \psi K^{(*)}$ \cite{eigen}  & $1.11 \pm 0.07 \pm 0.04$
\end{tabular}
\label{tab:fpfm}
\end{table}

\section{Acknowledgements}
We gratefully acknowledge the effort of the CESR staff in providing us with
excellent luminosity and running conditions.
M. Selen thanks the PFF program of the NSF and the Research Corporation, 
and A.H. Mahmood thanks the Texas Advanced Research Program.
This work was supported by the National Science Foundation, and the
U.S. Department of Energy.

% -------------------references

%\end{document}

%-----------------------------------------------------------------
%    Start all plots
%-----------------------------------------------------------------

%--------------figure 1----------mc only-----------------------------

%------------------figure 3 -----mombin leppi and leppi0-------------------

%------------------figure 4 ------------------------------

%------------------figure 5 ------------------------------

%------------------figure 6 ------------------------------

%------------------figure 7 ------------------------------

%------------------figure 8------------------------------

%------------------figure 9------------------------------

%------------------figure  10------------------------------

%------------------figure 11------------------------------

%------------------figure 12------------------ 

%-------------------figure 13----------------------------

%-----------------figure 14-----------leppi ws------------------------------

\end{document}